\documentstyle[12pt]{article}
\begin{document}
{\large\bf A built-in scale in the initial spectrum of density
perturbations: evidence from cluster and CMB data}
 \vskip 1cm 

\begin{center}
  {F. Atrio--Barandela}\\ 
{\it F{\'\i}sica Te\'orica, Facultad de Ciencias,
    Universidad de Salamanca, 37008 Spain\\ e--mail: atrio@astro.usal.es}\\ 

{J. Einasto}\\
{\it Tartu Observatory, EE-2444 T\~oravere, Estonia\\
 e--mail: einasto@max.aai.ee}\\

{S. Gottl\"ober, V. M\"uller}\\
 {\it Astrophysikalisches Institut Potsdam,  An
  der Sternwarte 16, D-14482 Potsdam, Germany\\ e--mail:
  (sgottloeber,vmueller)@aip.de}\\ 

{ A. Starobinsky}\\
{\it Landau Institute for Theoretical Physics, Moscow 117334, Russia \\
e--mail: alstar@landau.ac.ru}\\

\end{center}                           

\begin{abstract}
  
  We calculate temperature anisotropies of the cosmic microwave
  background (CMB) for several initial power spectra of density
  perturbations with a built-in scale suggested by recent optical
  data on the spatial distribution of rich clusters of galaxies.
  Using cosmological models with different values of spectral index,
  baryon fraction, Hubble constant and cosmological constant, we
  compare the calculated radiation power spectrum with the CMB
  temperature anisotropies measured by the Saskatoon experiment.  We
  show that spectra with a sharp peak at $120h^{-1}$Mpc are in agreement
  with the Saskatoon data. The combined evidence from cluster and CMB
  data favours the presence of a peak and a subsequent break in the
  initial matter power spectrum. Such feature is similar to the
  prediction of an inflationary model where an inflaton field is
  evolving through a kink in the potential.

\end{abstract}

\vskip 1cm

One of the crucial problems in cosmology is to determine the shape and
amplitude of the initial (primordial) power spectrum of density perturbations.
In the standard Friedman-Robertson-Walker cosmology this spectrum is
arbitrary. It is specified as an initial condition at the cosmological
singularity (the Big Bang). The only restriction is on the type of
perturbations: they should belong to those modes which do not destroy the
homogeneity of the Universe at early times, in particular, they should
represent the growing mode in the case of adiabatic perturbations. For the
scales of interest, the opposite assumption would result in the Universe being
strongly anisotropic and inhomogeneous at the time of the Big Bang
nucleosynthesis (BBN) that would completely spoil its predictions for the
primordial abundance of light elements.  On the other hand, the simplest
inflation models of the early Universe predict the power spectrum of the
growing mode of adiabatic perturbations at present to be approximately scale
invariant, i.e. Harrison-Zeldovich, characterised by a slope $n \approx 1$ on
large scales \cite{guth}.  In addition to processes during the inflation era
the current power spectrum is determined by physical processes occurring
during the radiation dominated regime that freeze out and damp the growth of
density perturbations within the cosmological horizon. The final spectrum
depends on the values of cosmological parameters and on the exact nature of
the dark matter present in the Universe.

>From the observational point of view, the current (evolved) power
spectrum of matter density perturbations can be estimated by measuring
clustering properties of galaxies and clusters of galaxies. Using the
distribution of rich Abell clusters, the spectrum has been recently
determined on scales from $k\approx 0.03$ up to $k\approx
0.3~h$~Mpc$^{-1}$ \cite{einasto} ($h$ is the Hubble constant in units
of $100~$km~s$^{-1}$Mpc$^{-1}$). The observed power spectrum contains
a sharp peak at $k\approx 0.05~h$~Mpc$^{-1}$. A similar feature on the same
scale has been observed in the 1-dimensional deep galaxy redshift
survey in the direction of Galactic poles, in the 2-dimensional power
spectrum obtained from the Las Campanas Redshift Survey of galaxies,
and from the deprojected power spectrum of the angular APM galaxy
survey \cite{landy}.

The purpose of this letter is to confront the power spectrum of matter density
perturbations obtained from cluster data with measurements of CMB temperature
anisotropies on different angular scales.  We shall concentrate on the
following questions: Are CMB data in agreement with the peaked matter power
spectrum? If so, can we find possible restrictions on the primordial initial
spectrum using {\it combined} CMB and optical data? In particular, can CMB and
optical data be explained by a specific choice of cosmological parameters
within the framework of the standard cosmological model with scale free
initial power spectrum, or  some change in this model is required?

In this letter, we do not intend to do full analysis of all available data at
intermediate and small angular scales on CMB.  We shall use the observations
made at Saskatoon \cite{saskatoon}.  By using a synthetic antenna beam, the
Saskatoon group was able to measure temperature anisotropies with five
different angular resolutions corresponding to multipoles between $l\approx
80$ and $l\approx 400$.  This range makes the experiment especially well
suited for comparison with the cluster power spectrum \cite{einasto} since it
roughly corresponds to the wavelengths probed by the cluster data. We used the
4-year COBE data \cite{bennet} to get the absolute normalisation and the shape
of the matter power spectrum at scales close to the present cosmological
horizon.

We calculate CMB temperature anisotropies for three different initial
power spectra: (a) a scale free initial spectrum with a power index
$n$, (b) a double power law approximation to the cluster spectrum, and
(c) a spectrum based on the observed cluster spectrum. Outside the
measured range, the latter was extrapolated assuming a scale free
spectrum.  At large wavenumbers ($k\ge 0.05$~$h$~Mpc$^{-1}$) the shape
of the observed cluster spectrum is similar to that of galaxies
\cite{costa}.  For the power spectrum (b), we used a slope $n=-1.8$
for small scales which is a smooth extrapolation of the cluster data.
However, since this region of the power spectrum has little influence
on multipoles above $l=400$, this assumption will not affect our
conclusions.  At large scales the spectrum is poorly determined.
Within observational errors, it is compatible with being
Harrison-Zeldovich. Furthermore, COBE/DMR data indicates \cite{bennet}
that the power spectrum of matter density perturbations has $n\approx
1$ for $k \approx 0.003~h$ Mpc$^{-1}$; more exactly, $n=1.1\pm 0.2$.
Accordingly, we varied the slope at large scales in that range.

\begin{figure*}[t]
\vspace*{17cm}
\caption{
  Goodness-of-fit contours of $\chi^2$ at 68\% and 95\% confidence level.  The
  $\chi^{2}$ statistics measures the deviation of the expected temperature
  anisotropy amplitude of a given model from the Saskatoon data.  The first
  row displays the results for the scale free model; the second row for the
  double power law model; and the lowest row for the cluster spectrum based
  model.  In the first column we plot models with varying Hubble constant and
  baryon fraction for a spectral index $n=1$ at large scales and no
  cosmological constant. In the middle column the same diagrams were repeated
  for $n=1.2$. Dashed lines indicate the nucleosynthesis bounds. The last
  column displays the results for models with different values of the
  cosmological constant. On all models with non-zero cosmological constant,
  the age of the Universe was chosen to be $14$ Gyr.}
\includegraphics{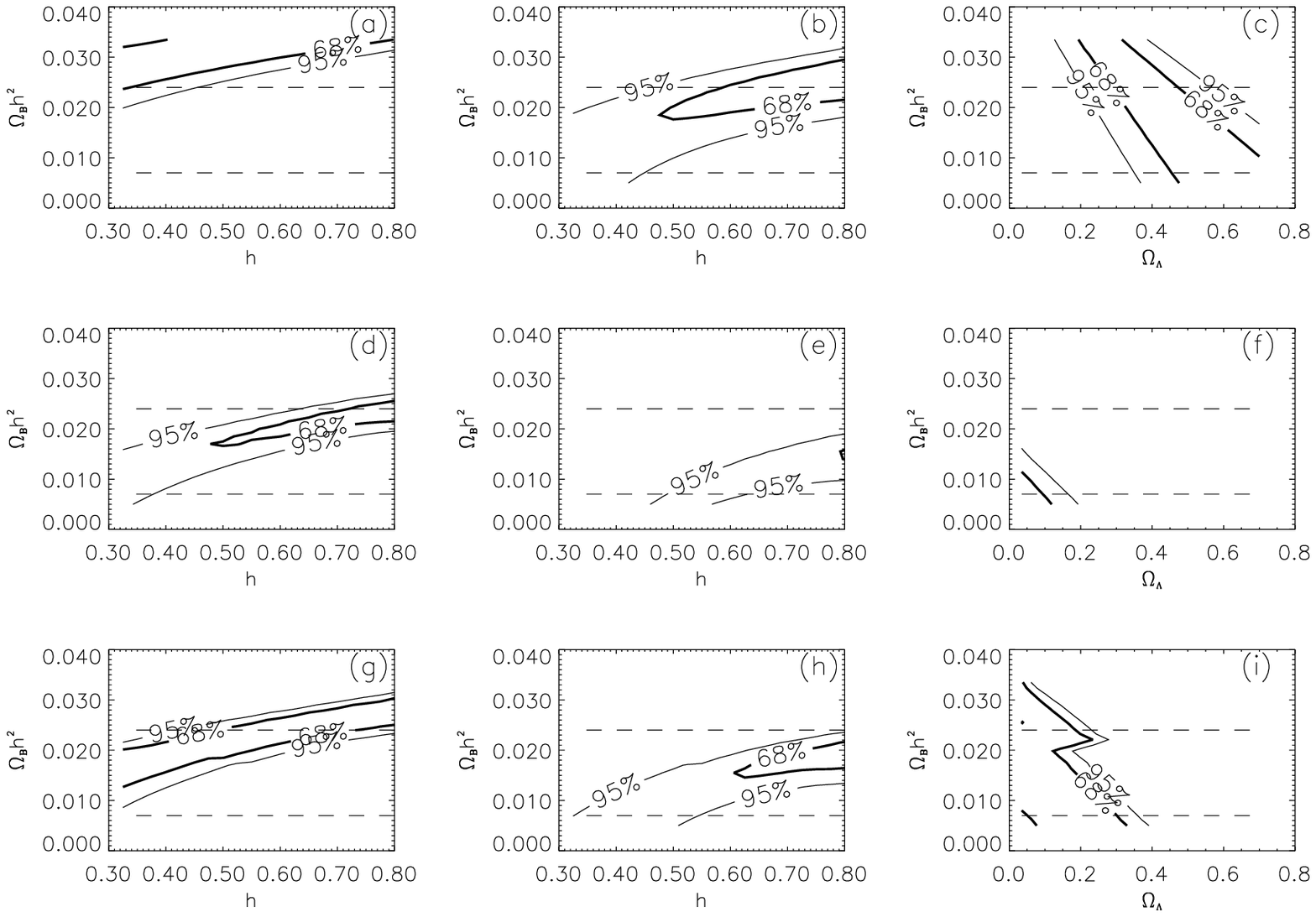}
\label{fig1}
\end{figure*}

The initial power spectrum is determined as follows:
\begin{equation}
P_{0}(k)=P(k)/T^{2}(k),
\label{init}
\end{equation} 
where $P(k)$ is the power spectrum of matter perturbations at the
current epoch, and $T(k)$ is the transfer function for a particular CDM
model. The transfer function depends only on physical processes taking place
within the horizon. In the previous expression we assumed that the observed
cluster power spectrum $P_{cl}(k)$ was proportional to $P(k)$ over the
range probed by the cluster data:
$P_{cl}(k)=b_{cl}^2P(k)$, where $b_{cl}$ is the bias factor for
clusters of galaxies.

In summary, for a set of cosmological parameters, we first calculate
the matter transfer function, and matter and radiation power
spectra for the scale invariant model (a); next, we determine the
initial power spectrum $P_{0}(k)$ for (b) and (c) as given by
eq.~(\ref{init}); finally, we calculate the angular CMB and matter
power spectra of models (b) and (c).  We assume the Universe has a
flat geometry.  We did not consider mixed dark matter (MDM) models
here.  They differ from CDM models mainly on small wavelengths which
have little influence on our results.  Furthermore, MDM models with
one stable neutrino and $h \ge 0.5$ have problems to create small
scale structure. In these models galaxy formation occurs too late
\cite{fang}, and considering scale-free MDM models with $n>1$ does not
help \cite{pog}.  In what follows we shall consider $\Omega_{b} +
\Omega_{c} + \Omega_{\Lambda} =1$, with $\Omega_{b}$, $\Omega_{c}$,
and $\Omega_{\Lambda}$ being the fraction of the energy density in
baryons, cold dark matter and vacuum energy (cosmological constant),
respectively.

To calculate the radiation power spectrum we used the packages COSMICS and
CMBFAST \cite{cmbfast}. The radiation power spectrum was normalised to the
COBE/DMR four year data \cite{bennet} using the angular wavenumber $C_{10}$ as
central value instead of the quadrupole \cite{white-bunn}. The comparison to
the cluster power spectrum gives $b_{cl}\approx 3$ for $n\approx 1$.  We have
performed the integration for the three primordial spectra and parameters:
$n=0.9, \dots, 1.3$; Hubble constant from $h=0.3$ to $0.8$; baryon density
from $\Omega_{b} h^{2}=0.005$ to $0.033$ centred on the range suggested by BBN
\cite{bbn}.  We also considered models with cosmological constant. In these
models we chose a Hubble constant that makes the Universe 14~Gyr old. It ranged
from $h=0.5$ for $\Omega_\Lambda = 0.1$ to $h=0.7$ for $\Omega_\Lambda = 0.7$,
in agreement with the recalibration of the Hubble constant and cosmic ages
made by \cite{feast} using the new determination of distances to subdwarfs and
Cepheids based on Hipparcos.  The amplitude of a temperature anisotropy
expected on a given angular scale was found using the window function that
best models the synthetic beam pattern of the Saskatoon experiment for that
scale \cite{saskatoon}.  Finally, for each model we calculated the
$\chi^2$-deviation between the theoretical prediction and the Saskatoon data.
In Figure~1 we plot the intervals in the parameter space at 68\% and 95\%
confidence level.  The dashed lines indicate the range $0.007 \le \Omega_b h^2
\le 0.024$ favoured by BBN \cite{bbn}.

Results of the $\chi^{2}$ test for comparison of different models and
cosmological parameters are shown in Figure~1. This figure shows that CMB data
alone exclude a large range of parameter space for each of our three basic
models. Standard CDM model is compatible with CMB data for values of the
Hubble constant and baryon fraction that are almost out of the range of
astronomical interest. An increase of the power index $n$ helps, but such
models are not viable because they overproduce mass fluctuations on scales of
$8 h^{-1}$~Mpc ($\sigma_8$). Actually, it is well known that even the
scale-invariant CDM model normalised to the COBE data is excluded for that
reason.  The best scale free model has a fairly large cosmological constant,
$\Omega_{\Lambda} \approx 0.6$.  A CDM model with large cosmological constant
and high baryon fraction was suggested in \cite{szalay} to explain the
presence of the large walls and voids in the distribution of galaxies.  Since
the baryon and dark matter fractions are comparable, the amplitude of acoustic
oscillations near the recombination epoch are fairly large. Below, we shall
discuss this model in more detail. On the other hand, the cluster based and
double power law spectra both fit the results of the Saskatoon experiment
rather well in the range of parameters of astronomical interest.  The best
agreement with the CMB data is obtained for a low or vanishing cosmological
constant, and for a spectral index $n=1$. The allowed range of the Hubble
constant is rather large for a reasonable baryon fraction.

In Figure~2 we compare matter power spectra and temperature anisotropy spectra
for our three basic models with the data.  The cosmological parameters were
chosen to reproduce the CMB data within the 68\% confidence level. As
expected, the temperature anisotropy spectra are very similar. In other words,
the present CMB data alone is not sufficient to discriminate between models.
By contrast, the matter power spectra are very different. The scale free model
with large cosmological constant has a broad maximum at large wavenumber
($k\approx 0.01~h~{\rm Mpc^{-1}}$); the maximum of the first acoustic
oscillation occurs at $k\approx 0.1~h~{\rm Mpc^{-1}}$, and is of rather small
relative amplitude.  Both wavenumbers are outside from the range where the
peak in the spectrum is observed: $k_{0}=0.052 \pm 0.005~h~{\rm Mpc^{-1}}$
\cite{einasto}.  Therefore we conclude that, contrary to the expectation of
\cite{szalay}, this peak is not related to acoustic oscillations in the
baryon--photon plasma. The scale free model spectrum agrees with the observed
cluster (and galaxy) spectrum on short wavelengths up to the peak.  However,
no combination of cosmological parameters reproduces the peak at $k=k_{0}$.
The existence of a broad maximum is an intrinsic property of all scale free
models and such maximum cannot produce a quasi-regular supercluster-void
network as shown in \cite{einasto2}. On the other hand, the cluster and double
power law spectra fit the observed cluster spectrum by construction and
reproduce the CMB data, i.e they fit equally well both datasets.  Therefore,
the present combined cluster and CMB data favour models with a built-in scale
in the {\it initial} spectrum.

Figure~2 shows also that differences between our three basic models are
on long wavelenghts for the matter power spectrum, and on small angular scales
for the CMB spectrum. Future measurements, both optical and CMB, should
concentrate to these scales.

\begin{figure*}[t]
\vspace*{8.5cm}
\caption{
  Comparison of matter power spectra and radiation temperature
  anisotropies with cluster and CMB data.  Dots with $1\sigma$ error
  bars give the observations: the measured cluster spectrum in the
  left panel and the Saskatoon data on CMB temperature anisotropies in
  the right panel. The scale free model spectra (short-dashed lines)
  were computed using the following parameters: $h=0.6$,
  $\Omega_{b}=0.07$, $\Omega_{c}=0.23$, and $\Omega_{\Lambda}=0.7$
  (short-dashed lines). This set of parameters produces  significant
  acoustic oscillations in the matter power spectrum within the
  parameter space allowed by Saskatoon. The cluster spectrum (solid
  lines) was calculated
  using $h=0.6$, $\Omega_{b}=0.08$, $\Omega_{c}=0.92$, and
  $\Omega_{\Lambda}=0$; and the double power law models (long-dashed) 
  using $h=0.6$, $\Omega_{b}=0.05$, $\Omega_{c}=0.95$, and
  $\Omega_{\Lambda}=0$.  To compare matter power spectra and
  observations we used a bias factor $b_{cl}\approx 3$.}
\includegraphics{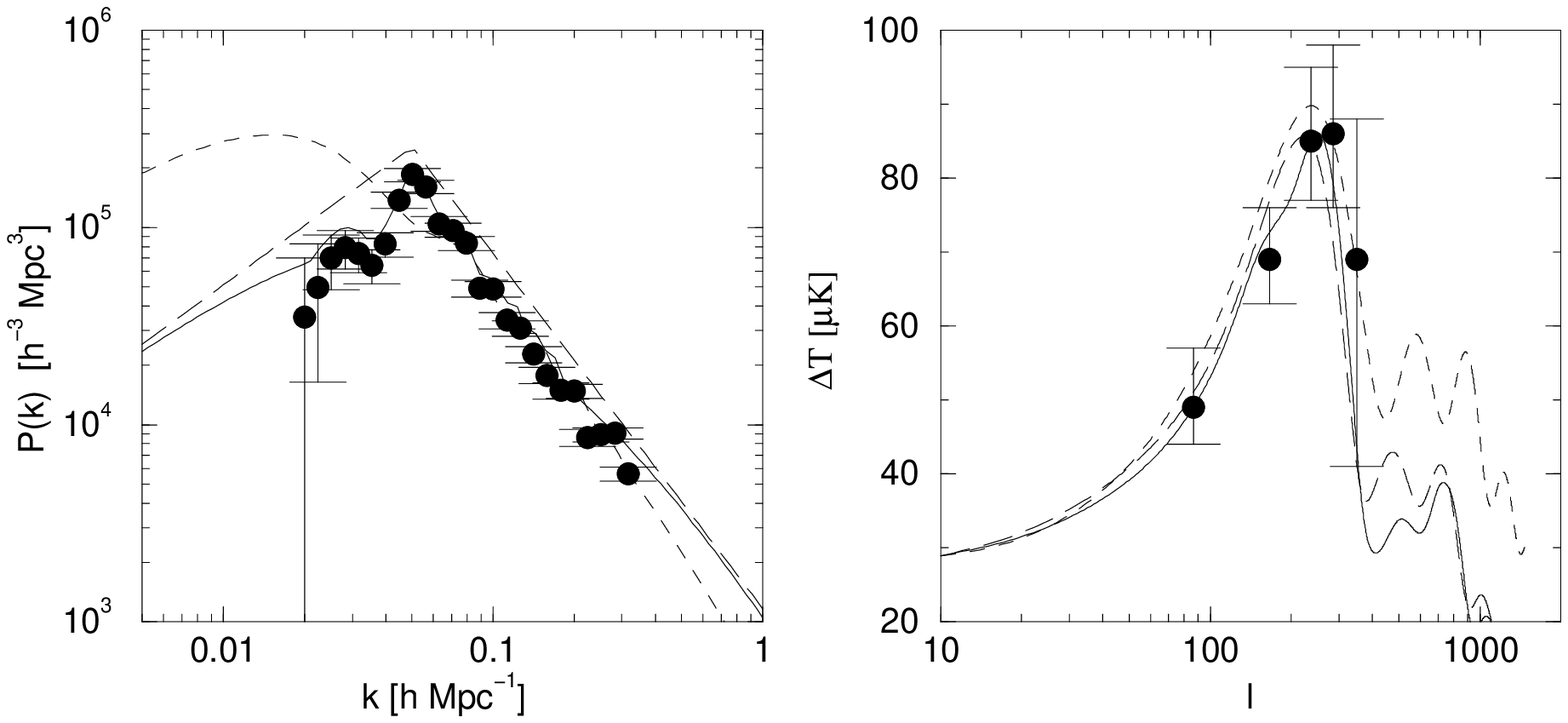}
\label{fig2}
\end{figure*}

Initial post-inflation power spectra are plotted in Figure~3. We see
that the cluster data based model has a peak at $k=0.05~h$~Mpc$^{-1}$,
followed by a break in amplitude.  The break in the initial spectrum
of the double power law model is similar to the previous one, but the
relative amplitude of the peak is larger.  Using both optical and CMB
data we could determine three parameters of the broken scale invariant initial
power spectrum: the position and the amplitude of the break, and
the amplitude of the peak above the initial scale free spectrum with
$n=1$.

\begin{figure*}[t]
\vspace*{8.5cm}
\caption{
  Initial power spectra. The left panel shows the initial power spectra
  for our three models. Parameters and line conventions are the same
  as in Figure~2.  The right panel shows a theoretical initial power
  spectrum generated in a model with an inflaton field evolving
  through a kink in the potential (solid line). The dashed line
  represents the scale free initial power spectrum. In both graphs the
  normalisation is arbitrary.}
\includegraphics{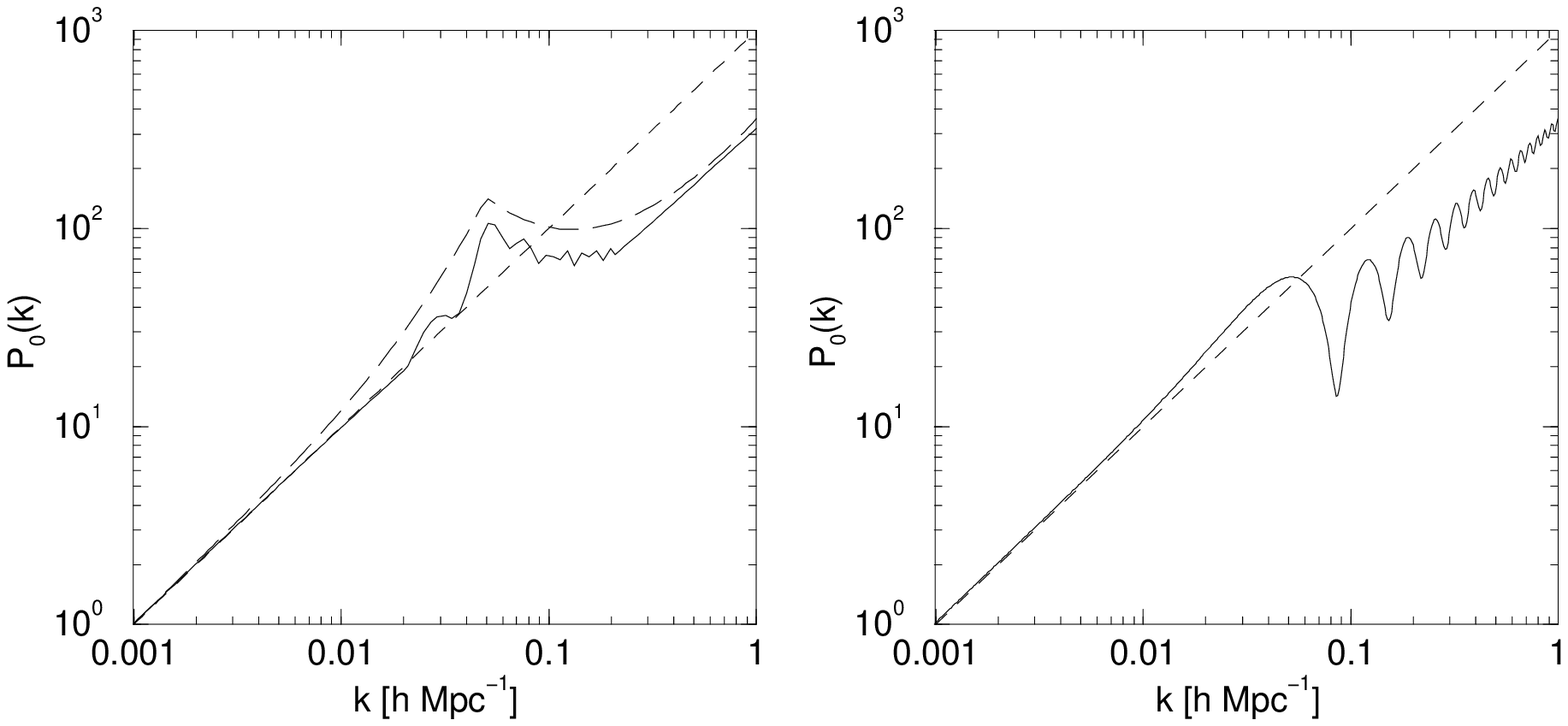}
\label{fig3}
\end{figure*}

Double inflation models provide a possible scenario where the formation of a
peak could have taken place \cite{kofman}. The presence of two scalar fields
driving the evolution of the Universe has a built-in scale defined by the
moment when the inflaton field that initially drove inflation becomes
subdominant.  Another possibility which does not require a fine tuning of the
initial energy density of different scalar fields is when the inflaton field
evolves through a kink in the potential.  A quick change in the first
derivative in the inflaton potential generates a sharp peak in the matter
power spectrum followed by a break in amplitude \cite{starobinski}.  Note that
this effect is beyond the slow-roll approximation to the motion of the
inflaton field or any adiabatic correction to it.  Obviously, the location of
this scale is a free parameter which can be determined from observational
data. The initial power spectrum found in \cite{starobinski} and calibrated
according the observed power spectrum is shown in Figure~3. It is very similar
to the empirical initial power spectrum plotted in the left panel of Figure~3;
the relative heights of the peak and the break are also close to the observed
values.

The main conclusion we can draw from our study is that, within the accuracy of
present measurements, the combined cluster and CMB temperature anisotropy data
suggest the existence of a break in the initial power spectrum of matter
density perturbations. The sharp peak found earlier in the cluster power
spectrum \cite{einasto} accounts for the observed high amplitude of the first
Doppler peak in the CMB spectrum, if the baryon fraction is not too high
($\Omega_bh^2 < 0.024$ for $h\simeq 0.6$).  On the other hand, if the
cosmological constant is large ($\Omega_{\Lambda}> 0.4$), then it would be
difficult to reconcile a built-in scale in the initial matter power spectrum
with the present CMB data.  Only new and more accurate observations of the
power spectrum, both optical and CMB, can discriminate between these two
alternatives.  In any case, the study of non-scale invariant initial spectra
are of crucial importance since they could provide a direct test of more
complicated models of inflation, violating the slow-roll approximation or
having more than one inflaton field.

\vskip 1cm {\bf Acknowledgements.} We thank Ed Bertschinger, Uros Seljak and
Matias Zaldarriaga for the permission to use their software packages COSMICS
and CMBFAST to calculate the angular power spectrum of the microwave
background radiation, and Alex Szalay for stimulating discussion.  This study
was supported by grants of the German Science Foundation, Estonian Science
Foundation, Astrophysical Institute Potsdam, and by Spanish German Integrated
Actions HA 1995 - 0079.  F.A.B. would like to acknowledge the support of the
Junta de Castilla y Le\'on, grant SA40/97. A.S. acknowledges the support of
the Russian Foundation for Basic Research, grant 96-02-17591. S.G.
acknowledges support from the Deutsche Akademie der Naturforscher Leopoldina
with means of the Bundesministerium f\"ur Bildung und Forschung grant LPD
1996.

\end{document}